\documentclass[aps,preprint]{revtex4}%
\usepackage{amsfonts}%
\usepackage{amsmath}%
\usepackage{amssymb}%
\usepackage{graphicx}
\begin{document}
\title{The Jungle Universe}
\author{J\'{e}r\^{o}me Perez}
\affiliation{ENSTA-Paristech
Applied Mathematics Laboratory
828, Boulevard des Marechaux, 91762 Palaiseau Cedex, France}
\author{Andr\'e F\"uzfa, Timot\'eo Carletti}
\affiliation{Namur Center for Complex Systems (naXys), University of Namur, Belgium}
\author{Laurence M\'elot}
\affiliation{Faculty of Computer Science, University of Namur, Belgium}
\author{Lazare Guedezounme}
\affiliation{UNESCO Chair of Mathematical Physics and Applications, University of Abomey-Calavi, Benin}
\begin{abstract}
In this paper, we exploit the fact that the dynamics of homogeneous and isotropic Friedmann-Lema\^{i}tre universes is a special case of generalized Lotka-Volterra system where the competitive species are the barotropic fluids filling the Universe. Without coupling between those fluids, Lotka-Volterra formulation offers a pedagogical and simple way to interpret usual Friedmann-Lema\^{i}tre cosmological dynamics.
A natural and physical coupling between cosmological fluids is proposed which preserve the structure of the dynamical equations.
Using the standard tools of Lotka-Volterra dynamics, we obtain the general Lyapunov function of the system when one of the fluids is coupled to dark energy. This provides in a rigorous form a generic asymptotic behavior for cosmic expansion in presence of coupled species, beyond the standard de Sitter, Einstein-de Sitter and Milne cosmologies. Finally, we conjecture that chaos can appear for at least four interacting fluids.
\end{abstract}
\maketitle

\section{Introduction}

Modern dynamical system theory can help us in understanding the evolution of
cosmological models. This remark in the introduction of the classical
Wainwright and Ellis textbook \cite{3}, is both true and full of sense : it is
often confronting approaches that one can really understand problems. In the
context of spatially homogeneous and anisotropic universes it was fully
investigated since the pioneering famous works of the 70's (e.g. \cite{bkl} for the BKL
conjecture, \cite{misner} for mixmaster universes or \cite{wain} for the
triangle map) up to mathematical proofs and deep understanding of this
dynamics by \cite{ringstrom} or extension to space-time dimensions $D>4$ e.g.
\cite{cosmobillard}.

The situation is a little bit less rich in the context of spatially
homogeneous and isotropic universes. Concerning the classical
Friedmann-Lema\^{\i}tre (FL) Universes containing non-interacting barotropic
fluids, various scenari for the fate of the Universe have been popularized. Among these are the Big Chill - when cosmic
expansion is endless or the Big Crunch - a final singularity of same nature than the
Big Bang for spatially closed cosmologies with vanishing or small cosmological
constant, or more recently Big Rip \cite{bigrip}, when Universe's scale factor
become infinite at a finite time in the future. Dynamical systems tools have
allowed some important results in the question of future asymptotic behavior
of cosmic expansion, for instance by demonstrating the existence of attracting
regimes and scaling solutions in quintessence models \cite{amendola2},
\cite{att}. Solutions to cosmological dynamics consists of time evolution of
density parameters associated to the barotropic fluids usually invoked to
model matter contents of the universe. The fate of the Universe is completely
related to its matter content. For example, Big Rip singularity occurs when
Universe contains the so-called \textquotedblright Phantom dark
energy\textquotedblright\ associated to a barotropic fluid with equation of
state $p=\omega\rho$ where the barotropic index $\omega<-1$. Recently, it
appears that cosmological FL models with \emph{interacting} components have
gained interest because it might be expected that the most abundant components in
the present Universe, dark energy (DE) and dark matter (DM), probably interact
with each other. Such interactions are considered by some authors to be
promising mechanisms to solve some of the $\Lambda$CDM problems like
coincidence (see for instance \cite{1}, \cite{2} and references therein). In
the literature the coupling between interacting fluids is generally
time-independant and quadratic in energy density $\mathcal{Q}_{ij}=\gamma
\rho_{i}\rho_{j}\ $(see \cite{4} and reference therein) where $\gamma$ is a
dimensioned constant or time-dependant and polynomial in energy density
$\mathcal{Q}_{ij}=\gamma H\rho_{i}^{m}\rho_{j}^{m-n}$ where $\ H$ is the
Hubble parameter and $m$ and $n$ relevant integers (see \cite{5} and
references therein).\ In these two categories of papers, new behaviors are
speculated for cosmological dynamics with non-linear interactions.\ In
particular the existence of cycles have been postulated if one of the species is barotropic with an
index $\omega<-1$.\ 

In this paper, we present for the first time FL Universe containing fluids in
interaction as a particular case of the well known
Lotka-Volterra system.\ This formulation is possible when one consider the
system in term of density contrasts $\Omega_{i}=$ $\frac{8\pi G}{3H^{2}}%
\rho_{i}$ evolving through the variable $\ln a$ where $a$ is the scale factor
of the FL universe.\ On the one hand this make us able to use a lot of
standard techniques of dynamical systems analysis in the context of
cosmology; and, on the other hand, this allows (as billiards do for
anisotropic models) a global comprehension of this important cosmological problem.

When there are no interactions between constitutive fluids, this formulation
allows to interpret those dynamics in a pedagogical way through an intuitive
and simple formulation. The FL cosmological dynamics can then be seen as a
competition between several species, each associated to one of the fluids
filling the universe. Those species all compete for feeding upon the same
resource which is spatial curvature. The usual asymptotic states of FL
dynamics, de Sitter, Einstein-de Sitter and Milne universes, can all be seen
as a particular equilibrium between cosmic species. This is the simplest
picture of the Jungle Universe.

In this paper we propose a new kind of coupling between fluids of the form
$\mathcal{Q}_{ij}=\gamma H^{-1}\rho_{i}\rho_{j}$.\ This is not an ad hoc approach since this the only way to preserve the natural Lotka-Volterra form of the FL
dynamics. Moreover this time-dependent coupling, once analyzed following
the scaling cosmology method presented by \cite{7}, shows that it grows with
the cosmic time.\ This last property makes this ansatz relevant both with
observational constraints and to avoid the coincidence problem.

The paper is structured as follow : in section \ref{section1} we show that FL
cosmological dynamics is actually a generalized Lotka-Volterra system; in
section \ref{section2} we interpret the FL cosmological dynamics in terms of
the generalized Lotka-Volterra system : The Jungle Universe; in section
\ref{section3}, we show how a general and physical interaction between
two fluids can preserve the structure of this dynamical system and we obtain
the general Lyapunov function for this kind of dynamics; in section
\ref{section4}, we generalize the formulation to $N$ directly coupled species
and focus in particular to triads ($N=3$) and quartets ($N=4$); finally, we
draw some conclusions in section \ref{section5}.

\subsection*{Notation}

In what follows, vectors are written bold faced (e.g. $\mathbf{r}\in
\mathbb{R}^{n}$) and the associated coordinates in the canonical basis are
denoted by the italic corresponding letters with an index (e.g. $\mathbf{r}%
=\left(  r_{1},\cdots,r_{n}\right)  ^{\top}$).

\section{Friedmann-Lema\^{\i}tre cosmology as generalized Lotka-Volterra
dynamical systems\label{section1}}

Taking into account a cosmological constant $\Lambda$, Einstein's equations of
general relativity write%
\[
R_{\mu\nu}-\frac{1}{2}g_{\mu\nu}R+\Lambda g_{\mu\nu}=\chi T_{\mu\nu}
\]
where $g_{\mu\nu}$ and $R_{\mu\nu}\ $are respectively the metric and the Ricci
tensors, $R$ is the scalar curvature (contraction of the Ricci), $T_{\mu\nu} $
is the stress-energy tensor and $\chi=8\pi Gc^{-4}$.\ The general paradigm of
standard cosmology consists of imposing
Friedmann-Lema\^{\i}tre-Robertson-Walker metric as an isotropic and
homogeneous description of the universe i.e.
\[
\mathrm{d}s^{2}=c^{2}\mathrm{d}t^{2}-a^{2}\left(  t\right)  \left[
\frac{\mathrm{d}r^{2}}{1-kr^{2}}+r^{2}\left(  \mathrm{d}\theta^{2}+\sin
^{2}\theta\mathrm{d}\phi^{2}\right)  \right]
\]
where $a\left(  t\right)  $ and $k$ are respectively the scale factor and the
curvature parameter, $t$ and $(r,\theta,\phi)$ being the synchronous time and
usual spherical coordinates, respectively .\ If one assumes that this universe
is filled by a perfect fluid of density $\rho$, pressure $p$ and
quadri-velocity field $u_{\mu}$ for which$\ T_{\mu\nu}=\left(  \rho
+c^{-2}p\right)  u_{\mu}u_{\nu}-pg_{\mu\nu}$, it is well known that the
dynamics of the universe are governed by Friedmann-Lema\^{\i}tre and
conservation equations :
\begin{align}
H^{2}  &  =\displaystyle\frac{8\pi G}{3}\rho+\displaystyle\frac{\Lambda c^{2}%
}{3}-\displaystyle\frac{kc^{2}}{3}\\
\displaystyle\frac{\ddot{a}}{a}  &  =-\displaystyle\frac{4\pi G}{3}\left(
\rho+3\frac{p}{c^{2}}\right)  +\displaystyle\frac{\Lambda c^{2}}{3}%
\label{acc}\\
\dot{\rho}  &  =-3H\displaystyle\left(  \rho+\frac{p}{c^{2}}\right)
\end{align}
where $H(t)=\displaystyle\frac{\dot{a}}{a}$ is the Hubble parameter and a dot
over a quantity indicates a derivation with respect to the synchronous time
$t$, the independent variable of the cosmological differential system. Both
parameters $k$ and $\Lambda$ might be seen as fixing the spatial and intrinsic
curvature of the geometry\footnote{If one interprets the cosmological constant
as the curvature associated to vacuum.}. Among the three above equations, only
two are independent since all are related through the second Bianchi
identities. The remaining two equations still include three unknown functions:
$\rho\left(  t\right)  $, $p\left(  t\right)  $ and $a\left(  t\right)  $.
This under-determination can be raised by introducing an equation of state for
the matter fluids.\ For example, barotropic fluids are such that $p=\omega
\rho$ where the constant $\omega\ $\ is called the barotropic index.\ In a
general physical way, this index ranges from $\omega_{\min}=-1 $ for scalar
field frozen in unstable vacuum to $\omega_{\max}=+1$ for stiff matter (e.g.
free scalar field) where sound velocity equals to speed of light. In this
paper we restrict our analysis to such barotropic physical fluids.
\newline\newline Following standard procedure, we rewrite the above equations
in terms of density parameters for matter $\Omega_{m}=\frac{8\pi G\rho}%
{3H^{2}}$, cosmological constant $\Omega_{\Lambda}=\frac{\Lambda}{3H^{2}}$,
curvature $\Omega_{k}=-\frac{k}{3a^{2}H^{2}}$ and deceleration parameter
$q=-\frac{\ddot{a}a}{\dot{a}^{2}}$. Friedmann-Lema\^{\i}tre equations and
energy conservation write for barotropic fluids therefore become
\[
\left\{
\begin{array}
[c]{l}%
1=\Omega_{m}+\Omega_{\Lambda}+\Omega_{k}\\
q=\frac{1}{2}\Omega_{m}\left(  1+3\omega\right)  -\Omega_{\Lambda}\\
\dot{\rho}=-3H\rho\left(  1+\omega\right)
\end{array}
\right.
\]
Please note that the latter equation can be directly integrated for constant
equation of state to give $\rho\sim a^{-3(1+\omega)}$.

Finally, we rewrite the above equations by changing the independent variable
to the number of efoldinds $\lambda=\log(a)$ and noting $^{\prime}$ for
$\lambda-$derivatives, one gets
\[
\left\{
\begin{array}
[c]{l}%
1=\Omega_{m}+\Omega_{\Lambda}+\Omega_{k}\\
\Omega_{m}^{\prime}=\Omega_{m}\left[  -\left(  1+3\omega\right)  +\left(
1+3\omega\right)  \Omega_{m}-2\Omega_{\Lambda}\right] \\
\Omega_{\Lambda}^{\prime}=\Omega_{\Lambda}\left[  2+\left(  1+3\omega\right)
\Omega_{m}-2\Omega_{\Lambda}\right]
\end{array}
\right.
\]
The dynamics of the Friedmann-Lema\^{\i}tre universe is contained in the two
last equations which form a differential system of generalized
Lotka-Volterra~\cite{Lotka1910,Volterra1926,Goel1971} equations well known in
population dynamics.\ As a matter or fact, introducing the dynamical vector
$\mathbf{x}=\left(  \Omega_{m},\Omega_{\Lambda}\right)  ^{\top}$and\ the
capacity vector $\mathbf{r}=\left(  -\left(  1+3\omega\right)  ,2\right)
^{\top}$, for $i=1,2\ $we have $x_{i}^{\prime}=x_{i}\ f\left(  x_{i}\right)  $
where the vector function $\ \mathbf{f}(\mathbf{x}) =\mathbf{r}+A\mathbf{x}$
is linear in the variables $x_{i}$, the community matrix $A$ being defined by
$\ $\
\[
\ \ \ A=\left[
\begin{array}
[c]{cc}%
\left(  1+3\omega\right)  & -2\\
\left(  1+3\omega\right)  & -2
\end{array}
\right]
\]
This formulation allows us to assimilate the dynamics of
Friedmann-Lema\^{\i}tre universes to those of a competition between species,
represented by $\Omega_{m}\ $and$\ \Omega_{\Lambda}$, for the resources in
$\Omega_{k}$. This point of view is not anecdotal and will reveal a lot of
benefit : such equations are very well known to the dynamical system
specialist, it allows a lot of intuitive non trivial results, establish an
analogy that will help us deriving new cosmological behavior for coupled
models besides of providing a pedagogic and interesting insight on cosmic expansion.

First of all, it is easy to see that orbits cannot cross the $\Omega_{m}=0$ or
$\Omega_{\Lambda}=0\ $axes which are orbits themselves. As the matrix $A$
fully degenerates (rank equal to 1)it is clearly not invertible, equilibrium
points must lie on axis. In particular as denoted by \cite{UL} or
\cite{Hobson} using a slightly different dynamical system, there exists 3
equilibria which are Milne universe $\mathbf{x}_{0}=\left(  0,0\right)  $,
Einstein-de Sitter universe $\mathbf{x}_{1}=\left(  1,0\right)  $ and de
Sitter universe $\mathbf{x}_{2}=\left(  0,1\right)  $. Using the large
knowledge of such systems from bio-mathematics (e.g. \cite{HS},\cite{Murray})
the $\mathbf{r}$\ vector contains the intrinsic birth or death rates of the
species.\ The dynamics of competitive Lotka-Volterra systems with such a
degenerate matrix is well known:

\begin{itemize}
\item If the initial condition is located in the positive quadrant
$Q^{+}=\left\{  \Omega_{m}>0\right\}  \times\left\{  \Omega_{\Lambda
}>0\right\}  $ then $\mathbf{x}\rightarrow\mathbf{x}_{2}$ when $t$ or
$\lambda$ goes to infinity, the reason of this attractive character of the de
Sitter universe is uniquely contained in the fact that $r_{2}\geq r_{1}$ for
all physical values of the barotropic index $\omega$. If we extend values of
$\omega$ considering phantom dark energy instead of pressureless matter by
letting $\omega<-1$ the attractor become the (phantom DE-dominated)
Einstein-de Sitter universe ($\mathbf{x}_{1}$) simply because in this case
$r_{1}\geq r_{2}$. This is obvious since in this case the energy density of
the phantom DE grows like a power-law with the scale factor ($\rho_{DE}\sim
a^{-3(1+\omega)}$ where $\omega<-1$), therefore asymptotically dominating the
constant density associated to the cosmological term.

\item If the initial condition lies on the $\Omega_{m}$ axis the attractor is
the Einstein-de Sitter universe if $\omega<-\frac{1}{3}$ and Milne universe
($\mathbf{x}_{0}$) if $\omega\geq-\frac{1}{3}$. This is obvious since, in the
absence of a cosmological constant ($\Omega_{\Lambda}=0$), the competition is
left between matter and curvature energy densities, the latter decreasing as
$a^{-2}$. Therefore, asymptotic dominance of matter is only possible when
$\omega<-1/3$, so that the related density can eventually dominate (since it
scales as $\rho_{m}\sim a^{-3(1+\omega)}$).

\item If the initial condition lies on the $\Omega_{\Lambda}$ axis the
attractor is the de Sitter universe for any values of $\omega.$ Once again,
this is obvious since asymptotically the constant energy density of the
cosmological term will dominate the decreasing energy density related to the curvature.
\end{itemize}

These results are well known and presented in a slightly different manner in
\cite{UL} or \cite{Hobson}. The new point here is the dynamical population
formulation of the problem and interesting results will be derived through
usual techniques in dynamical system theory. We will also present new
cosmological consequences on coupled models which are directly inspired by the
analogy with evolution of populations in competition. One possibility consists
of investigating how far the natural cyclic orbits appearing usually in
population dynamics could appear in standard cosmology.\ This is the object of
the next section.

\section{Multi-components Friedman-Lema\^{\i}tre Universes : Jungle
Universes.\label{section2}}

In the latter section we have presented the generalized Lokta-Volterra
formulation for the dynamics of usual Friedmann-Lema\^{\i}tre universe with
non-vanishing cosmological constant. In particular we have only considered one
simple barotropic fluid characterized by a given value of $\omega$.\ We can
generalize this situation to the more complicated yet realistic case where the
universe is filled by several kinds of barotropic fluids without any direct
interactions.\ In this section, we consider for example baryonic matter
(b$-$indiced and for which $\omega_{\mathrm{b}}=0$) and radiation (r$-$indiced
and for which $\omega_{\mathrm{r}}=\frac{1}{3}$).\ It is well known that the
repulsive feature obtained with a positive cosmological constant can also
advantageously be obtained through some dark energy fluid component
(e$-$indiced) associated to a barotropic index $\omega_{\mathrm{e}}\in\left[
-1,-1/3\right]  ;$ the cosmological constant term could then be obtained
taking $\omega_{\mathrm{e}}=-1$. In the following, roman indexes refer to the
fluid component considered.

The cosmological term in Friedmann-Lema\^{\i}tre equations can therefore be
removed, introducing the densities $\Omega_{\mathrm{x}}=\frac{8\pi
G\rho_{\mathrm{x}}}{3H^{2}}$ for $\mathrm{x}=\mathrm{b},\mathrm{r}$ and
$\mathrm{e}$ including the conservation of each kind of fluids they write%
\[%
\begin{array}
[c]{l}%
1=\Omega_{\mathrm{b}}+\Omega_{\mathrm{r}}+\Omega_{\mathrm{e}}+\Omega
_{\mathrm{k}}\\
2q=\Omega_{\mathrm{b}}+2\Omega_{\mathrm{r}}+\left(  1+3\omega_{\mathrm{e}%
}\right)  \Omega_{\mathrm{e}}\\
\left(  \ln\rho_{\mathrm{x}}\right)  ^{\prime}=-3(1+\omega_{\mathrm{x}%
)}\ \ \mathrm{for}\ \mathrm{x}=\mathrm{b},\mathrm{r\ and}\ \mathrm{e;}%
\end{array}
\]
A basic calculus shows that $\left(  \ln H\right)  ^{\prime}=-q-1$ hence
Friedmann-Lema\^{\i}tre equations write
\[
\frac{\Omega_{\mathrm{x}}^{\prime}}{\Omega_{\mathrm{x}}}=\left(  \ln
\Omega_{\mathrm{x}}\right)  ^{\prime}=\Omega_{\mathrm{b}}+2\Omega_{\mathrm{r}%
}+\left(  1+3\omega_{\mathrm{e}}\right)  \Omega_{\mathrm{e}}-3\omega
_{\mathrm{x}}-1 \quad\mathrm{for}\; \mathrm{x}=\mathrm{b},\mathrm{r\ and}\;
\mathrm{e}
\]
The three dimensional differential system for $\Omega_{\mathrm{e}}$,
$\Omega_{\mathrm{b}}$ and $\Omega_{\mathrm{r}}$ is always a generalized
Lotka-Volterra form with a fully degenerate community matrix. The dynamics is
then always governed by the capacity vector\ $\mathbf{r=}\left[
-1,-2,-3\omega_{\mathrm{e}}-1\right]  $ which actually rules the asymptotic
behavior. Besides of the origin, there is now one additional equilibrium on
each axis and if $\mathbf{r}$ possesses a component which is greater than all
others, the corresponding equilibrium with this component maximal is globally
stable over the positive orthant. This smart result is sufficient to claim
that dark energy (for which $\omega_{\mathrm{e}}\in\left[  -1,-1/3\right]  $)
correspond to this $\mathbf{r}$ maximal components and then the universe such
that $\Omega_{\mathrm{b}}=\Omega_{\mathrm{r}}=0$ and $\Omega_{\mathrm{e}}=1$
is globally stable out from axis $\Omega_{\mathrm{b}}=0$ and $\Omega
_{\mathrm{r}}=0$.

This three dimensional situation is readily generalizable to any number of non
interacting fluids each governed by a separated conservation equation. \ The
dynamical behavior is asymptotically always the same : the system evolves like
a competitive one in which all species (predators) are fed by the same prey
(which is curvature...).
Asymptotically and out of axis, only one species survives, the one which
possesses the greater value of $-3\omega_{\mathrm{x}}-1$. This species is
always the dark energy fluid in our physical hypotheses $\omega\in\left[
-1,1\right]  $. Once the Universe is filled with even a small amount of dark
energy, there is no way it cannot dominate forever the fate of the cosmos.
This is Jungle Law for a Jungle Universe. Fortunately, this will cease to be
true, as we shall see in the next section, if dark energy is not so dark, but
it is in interaction with the other components.

\section{Cooperation in the Jungle Universes\label{section3}}

\subsection{General dynamics with dark coupling}

In the last sections we have presented a way to express the dynamics of
Friedmann-Lema\^{\i}tre universes using generalized Lotka-Volterra
differential system theory. This also offers new perspectives in determining
cosmological analogues of specific cases in competitive dynamics. It is well
known that the generic dynamics of such systems contains limit cycles or
periodic orbits. We will describe in this section how direct coupling can be
used to bring such a behavior in the context of cosmology.

When the fluids filling the universe are not interacting with each other, the
community matrix of the generalized Lotka-Volterra system must have the same
rows and then must be fully degenerated. In order to make its rank greater
than one, we must introduce coupling between species, i.e. interactions
between cosmological fluids. On the other hand, this kind of interactions is
broadly used in cosmology, with the coupling between inflaton and radiation
during reheating (e.g., \cite{martin}) or the one between dark matter and dark
energy (e.g., \cite{amendola2,caldera,amendola}), or even the decay of heavy
matter particles like WIMPS into light relativistic particles (e.g.,
\cite{DMdecay}). Modern cosmology make strong use of coupled fluids for a
variety of purposes, therefore making this study of coupled models in terms of
Lotka-Volterra systems of first heuristic interest.

In order to show the phenomenon we will present in this section the situation
where the universe contains radiation, baryonic matter, dark matter
($\mathrm{d}-$indiced)\footnote{Although both are pressureless with $\omega
=0$, we split both to allow for different couplings.}, dark energy and we
suppose a coupling between the two dark components. This constitutes a coupled
quintessence scenario \cite{amendola}. On one hand, it is necessary to
preserve the global energy conservation as imposed by Noether theorem and
Poincare invariance, energy transfer must compensate in the global energy
balance. Hence, at each time, the part of the energy taken by the first
component must be given to the other to which it couples. To achieve this,
conservation equations for two coupled dark fluids must be of the following form:%

\[
\left\{
\begin{array}
[c]{l}%
\dot{\rho}_{\mathrm{d}}=-3H\rho_{\mathrm{d}}\left(  1+\omega_{\mathrm{d}%
}\right)  +\mathcal{Q}\ \\
\dot{\rho}_{\mathrm{e}}=-3H\rho_{\mathrm{e}}\left(  1+\omega_{\mathrm{e}%
}\right)  -\mathcal{Q}%
\end{array}
\right.
\]
where $\mathcal{Q}$ represents the energy transfer. This coupling leaves
unchanged the global energy-momentum conservation, it is then invisible in
standard general relativity and it glimpses at (micro-)physics describing dark
components of the universe.\ In literature, one usually finds that this energy
transfer is arbitrarily expressed as a linear combination of the dark sector
densities:
\[
\mathcal{Q}=A_{\mathrm{d}}\rho_{\mathrm{d}}+A_{e}\rho_{\mathrm{e}}%
\]
where the coefficients are either proportional to Hubble parameter $H$ either
constant (see \cite{amendola2,caldera,amendola}). In this paper, we introduce
a new non-linear parametrization of the energy transfer that allows us
matching the coupled model to a general Lotka-Volterra system. This ansatz is
given by
\begin{equation}
\mathcal{Q}=\frac{8\pi G}{3H}\varepsilon\rho_{\mathrm{e}}\rho_{\mathrm{d}}%
\end{equation}
where the \emph{coupling parameter }$\varepsilon$\emph{ is a positive
constant}. This form of coupling is needed to preserve the fundamental
generalized Lotka-Volterra form of the FL dynamical system, but it have also
good physical properties. As a matter of fact, if we consider only the dark
components of the universe $-$ the dark plane where $\Omega_{\mathrm{b}%
}=\Omega_{\mathrm{r}}=0$ $-$ in the case where the non baryonic dark matter is
non-relativistic and pressureless, i.e. $\omega_{\mathrm{d}}=0$ and following
the scaling method introduced by \cite{7}, we get
\[
\frac{d}{dt}\left(  \frac{\rho_{\mathrm{d}}}{\rho_{\mathrm{e}}}\right)
=\frac{\rho_{\mathrm{d}}}{\rho_{\mathrm{e}}}\left[  \frac{\dot{\rho
}_{\mathrm{d}}}{\rho_{\mathrm{d}}}-\frac{\dot{\rho}_{\mathrm{e}}}%
{\rho_{\mathrm{e}}}\right]  =\frac{\rho_{\mathrm{d}}}{\rho_{\mathrm{e}}%
}\left[  3H\omega_{\mathrm{e}}+\frac{8\pi G\varepsilon\left(  \rho
_{\mathrm{d}}+\rho_{\mathrm{e}}\right)  }{3H}\right]
\]
In the dark plane we have
\[
\frac{8\pi G}{3}\left(  \rho_{\mathrm{d}}+\rho_{\mathrm{e}}\right)  =H^{2}%
\]
as $H=\frac{d\ln a}{dt}$ we obtain
\[
\frac{d}{dt}\left(  \frac{\rho_{\mathrm{d}}}{\rho_{\mathrm{e}}}\right)
=\frac{\rho_{\mathrm{d}}}{\rho_{\mathrm{e}}}H\left[  3\omega_{\mathrm{e}%
}+\varepsilon\right]  \ \ \Longrightarrow\ \frac{d\ln\left(  \frac
{\rho_{\mathrm{d}}}{\rho_{\mathrm{e}}}\right)  }{dt}=\left[  3\omega
_{\mathrm{e}}+\varepsilon\right]  \frac{d\ln a}{dt}%
\]
and finally
\begin{equation}
\frac{\rho_{\mathrm{d}}}{\rho_{\mathrm{e}}}=\frac{\rho_{\mathrm{d},0}}%
{\rho_{\mathrm{e},0}}\left(  \frac{a}{a_{0}}\right)  ^{3\omega_{\mathrm{e}%
}+\varepsilon}\ \ \ \label{coincidence}%
\end{equation}
which avoid the coincidence problem if $3\omega_{\mathrm{e}}+\varepsilon>0$.

Since the Raychaudhuri equation (\ref{acc}) and consequently $\left(  \ln
H\right)  ^{\prime}$ are left unchanged by the introduction of such
couplings \footnote{This is so since gravity is still minimally coupled to
matter fluids.}, but we have now%

\[%
\begin{array}
[c]{lll}%
\left(  \ln\Omega_{\mathrm{d}}\right)  ^{\prime} & = & \left(  \ln
\rho_{\mathrm{d}}\right)  ^{\prime}+2q+2\\
& = & \Omega_{\mathrm{b}}+\left(  1+3\omega_{\mathrm{d}}\right)
\Omega_{\mathrm{d}}+2\Omega_{\mathrm{r}}+\left(  \varepsilon+1+3\omega
_{\mathrm{e}}\right)  \Omega_{\mathrm{e}}-\left(  3\omega_{\mathrm{d}%
}+1\right)
\end{array}
\]

and%
\[%
\begin{array}
[c]{lll}%
\left(  \ln\Omega_{\mathrm{e}}\right)  ^{\prime} & = & \left(  \ln
\rho_{\mathrm{e}}\right)  ^{\prime}+2q+2\\
& = & \Omega_{\mathrm{b}}+\left(  1+3\omega_{\mathrm{d}}-\varepsilon\right)
\Omega_{\mathrm{d}}+2\Omega_{\mathrm{r}}+\left(  1+3\Omega_{\mathrm{d}%
}\right)  \Omega_{\mathrm{d}}-\left(  3\omega_{\mathrm{e}}+1\right)  \,.
\end{array}
\]
The other two remaining equations for $\Omega_{\mathrm{b}}$ and $\Omega
_{\mathrm{d}}$ are not affected by the dark coupling. With this coupling and
under these last hypotheses the generalized Lotka-Volterra equations
associated to isotropic, homogeneous and barotropic fluid filled \ universe
for the dynamical variable $\mathbf{x}=\left(  \Omega_{\mathrm{b}}%
,\Omega_{\mathrm{d}},\Omega_{\mathrm{r}},\Omega_{\mathrm{e}}\right)  ^{\top}$
are defined by a capacity vector $\mathbf{r}$ and a community matrix $A$ such
that
\begin{equation}
A=\left[
\begin{array}
[c]{cccc}%
1 & 1 & 2 & 1+3\omega_{\mathrm{e}}\\
1 & 1 & 2 & \varepsilon+1+3\omega_{\mathrm{e}}\\
1 & 1 & 2 & 1+3\omega_{\mathrm{e}}\\
1 & 1-\varepsilon & 2 & 1+3\omega_{\mathrm{e}}%
\end{array}
\right]  \quad\mathrm{and}\;\mathbf{r}=\left[
\begin{array}
[c]{c}%
-1\\
-1\\
-2\\
-1-3\omega_{\mathrm{e}}%
\end{array}
\right]  \label{sysdyn1}%
\end{equation}
As desired this matrix is not fully degenerate but has generally its rank
equal to 3. This dynamic is characterized by five equilibria in the positive
quadrant which are $\mathbf{\tilde{x}}^{0}=\left(  0,0,0,0\right)  ^{\top}$,
$\mathbf{\tilde{x}}^{1}=\left(  0,0,1,0\right)  ^{\top}$, $\mathbf{\tilde{x}%
}^{2}=\left(  0,0,0,1\right)  ^{\top}$, $\mathbf{\tilde{x}}^{3}=\left(
1-\alpha,\alpha,0,0\right)  ^{\top}$ with $\alpha\in\left]  0,1\right]  $ and
$\mathbf{\tilde{x}}^{4}=\varepsilon^{-1}\left(  0,-1-3\omega_{\mathrm{e}%
},0,1\right)  ^{\top}$ the first four being globally unstable while the last
is by far the most interesting.

Provided that
\begin{equation}
\omega_{\mathrm{e}}<-\frac{1}{3}\mbox{ and }\varepsilon>-3\omega_{\mathrm{e}%
}>0 \label{wecond}%
\end{equation}

\bigskip the equilibrium $\mathbf{\tilde{x}}^{4}$ is always in the positive
quadrant but it is no more hyperbolic and two complex eigenvalues, namely
\[
\lambda_{\pm}=\pm i\sqrt{\frac{\left\vert 9\omega_{\mathrm{e}}^{2}%
+3\varepsilon\omega_{\mathrm{e}}+3\omega_{\mathrm{e}}+\varepsilon\right\vert
}{\varepsilon}},
\]
occur in the spectrum of the linearized dynamics around $\mathbf{\tilde{x}%
}^{4}$.\ A precise analysis of the dynamical behavior of the system is then
required. In order to do this we have decomposed the job into two parts :

\begin{enumerate}
\item In a first step (section \ref{sect:cycle}), we have restricted the
analysis to the dark plane $\left(  \Omega_{\mathrm{d}},\Omega_{\mathrm{e}%
}\right)  $ where we have rigorously proven that the dynamic is generally
cyclic; this proof was exhibited using a general new Lyapunov function.

\item In a second step (section \ref{sect:att}), we have shown that this dark
plane is attractive for all orbits whose initial conditions belong to the
hyper-tetrahedron
\begin{equation}
T_{4}=\{\Omega_{\mathrm{b}}>0\}\cup\{\Omega_{\mathrm{d}}>0\}\cup
\{\Omega_{\mathrm{r}}>0\}\cup\{\Omega_{\mathrm{e}}>0\}\cup\{\Omega
_{\mathrm{b}}+\Omega_{\mathrm{d}}+\Omega_{\mathrm{r}}+\Omega_{\mathrm{e}%
}<1\}\,. \label{eq:htetra}%
\end{equation}

\end{enumerate}

\subsection{\bigskip Cyclicity of orbits in the dark plane : Lyapunov function
for FL\ dynamics\label{sect:cycle}}

Cyclic behaviors for coupled FL cosmologies has been proposed by some papers
(see \cite{4} or \cite{5}) but in each cases the polytropic index
$\omega_{\mathrm{e}}$ required for dark energy is less than $-1$. Moreover
their conclusion of cyclic orbits are obtained from linear analysis around non
hyperbolic equilibria; however, it is well known in dynamical system analysis
that one cannot conclude anything in this case: we have devoted the appendix \ref{appnonhyp}
to this point. In order to be sure to have such cyclic behavior it is
necessary in this context to use more refined tools like Lyapunov functions.

Such tools are fundamentals in physics but generally it is very hard to find
them.However, it is possible in a very general manner in context of our
coupled FL dynamics thanks to the fact that it is a generalized Lotka-Volterra system.

In an pedagogical objective we propose to show how to construct such kind of
fundamental functions for Jungle Universes containing a coupling in the dark
sector. Using this method, this result is widely generalizable to another cases.

In the so-called dark-plane $\left(  \Omega_{\mathrm{b}}=\Omega_{\mathrm{r}%
}=0\right)  $ and with the notations $x=\Omega_{\mathrm{d}}$ and
$y=\Omega_{\mathrm{e}}$, the dynamics is then governed by the generalized
Lotka-Volterra system
\[
\left\{
\begin{array}
[c]{l}%
x^{\prime}=x\left[  x+\left(  1+3\omega_{\mathrm{e}}+\varepsilon\right)
y-1\right] \\
y^{\prime}=y\left[  \left(  1-\varepsilon\right)  x-\left(  1+3\omega
_{\mathrm{e}}\right)  y-1-3\omega_{\mathrm{e}}\right]
\end{array}
\right.
\]
hence
\[
\mathbf{r}_{2}=\left[
\begin{array}
[c]{c}%
-1\\
-1-3\omega_{\mathrm{e}}%
\end{array}
\right]  \ \ \ \mathrm{and}\ \ \ A_{2}=\left[
\begin{array}
[c]{cc}%
1 & 1+3\omega_{\mathrm{e}}+\varepsilon\\
1-\varepsilon & 1+3\omega_{\mathrm{e}}%
\end{array}
\right]
\]
As indicated in the previous section,for the dark sector and under the
restriction $\left(  \ref{wecond}\right)  ,$ there is a unique equilibrium in
the strict positive quadrant, namely $\left(  \tilde{x},\tilde{y}\right)
=\varepsilon^{-1}\left(  -1-3\omega_{\mathrm{e}},1\right)  $. The interval of
study $\omega_{\mathrm{e}}\in\left(  -\infty,-\frac{1}{3}\right]  $ include
the cosmological constant ($\omega_{\mathrm{e}}=-1$), a large variety of
quintessence scenario, and of course the hypothetical phantom dark matter such
that $\omega_{\mathrm{e}}<-1$. \ In all these cases, non hyperbolic
eigenvalues appear for a sufficient strength of coupling $\varepsilon
>\left\vert 3\omega_{\mathrm{e}}\right\vert $, for lighter coupling
$\mathbf{\tilde{x}}^{4}$ is unstable.

Let us turn now to the construction of the Lyapunov function.

Using a bit of intuition and dynamical population analysis tools (e.g.
\cite{HS}) one can use the function $V_{\varepsilon,\omega_{\mathrm{e}}%
}\left(  x,y\right)  =x^{\alpha}y^{\beta}\left(  a+bx+cy\right)  $ where
$\alpha$ and $\beta$ are functions of $\varepsilon$ and $\omega_{\mathrm{e}}$;
$a,b$ and $c$ are three constants, all being determined in order to obtain a
Lyapunov function. As $A_{2}$ is now invertible choosing $\left(  \alpha
,\beta\right)  ^{\top}=A_{2}^{-\top}\mathbf{r}_{2}$ i.e. $\alpha
=-\frac{1+3\omega_{\mathrm{e}}}{\varepsilon+3\omega_{\mathrm{e}}}$ and
$\beta=\frac{1}{\varepsilon+3\omega_{\mathrm{e}}}$, it is easy to check that
\[
V_{\varepsilon}^{\prime}=x^{\alpha}y^{\beta}\left[  \left(  b-c\right)
\varepsilon xy-\left(  3\omega_{\mathrm{e}}a+c+3c\omega_{\mathrm{e}}+a\right)
y-\left(  a+b\right)  x\right]
\]
Hence, choosing finally $a=-c$, $b=c$ we can construct the function
\[
L_{\varepsilon,\omega}\left(  x,y\right)  =\kappa V_{\varepsilon
,\omega_{\mathrm{e}}}\left(  x,y\right)  -1
\]
where
\[
V_{\varepsilon,\omega_{\mathrm{e}}}\left(  x,y\right)  =x^{-\frac
{1+3\omega_{\mathrm{e}}}{\varepsilon+3\omega_{\mathrm{e}}}}y^{\frac
{1}{\varepsilon+3\omega_{\mathrm{e}}}}\left(  x+y-1\right)  \mbox{ and }\kappa
^{-1}=V_{\varepsilon,\omega_{\mathrm{e}}}\left(  \tilde{x},\tilde{y}\right)
\]
one can verify that

\begin{enumerate}
\item $L_{\varepsilon,\omega_{\mathrm{e}}}\left(  \tilde{x},\tilde{y}\right)
=0$,

\item $L_{\varepsilon,\omega_{\mathrm{e}}}\left(  x,y\right)  >0$ if $\left(
x,y\right)  \in\mathbb{R}^{2}\setminus\left(  \tilde{x},\tilde{y}\right)  $

\item $L_{\varepsilon,\omega_{\mathrm{e}}}^{\prime}\left(  x,y\right)  =0$
\mbox{ for all values of } $\left(  x,y\right)  \in\mathbb{R}^{2}$
\end{enumerate}

\begin{figure}[th]
\begin{center}
\includegraphics[scale=1]{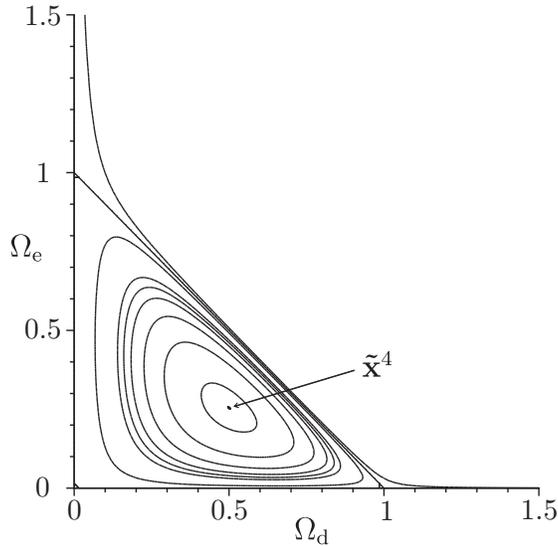}
\end{center}
\caption{Contour levels of $L_{4,-1}\left(  x,y\right)  $}%
\label{fig_contourseps4}%
\end{figure}

Hence the function $L\left(  x,y\right)  _{\varepsilon,\omega_{\mathrm{e}}}$
is a Lyapunov function for this dynamics and orbits are confined on curves
$L_{\varepsilon,\omega_{\mathrm{e}}}\left(  x,y\right)  =\mu$ where $\mu$ is
any positive constant. Such curves are plotted on figure
\ref{fig_contourseps4} for the generic values $\varepsilon=4$ and
$\omega_{\mathrm{e}}=-1$.

It must be noted that when $\Omega_{\mathrm{d}}+\Omega_{\mathrm{e}}<1$, the
dynamics in the dark plane is periodic as all the contour levels of
$V_{\varepsilon}$ are closed and all solutions are maximal. The corresponding
cosmological solution correspond to endless oscillations of the density
parameters $\left(  \Omega_{\mathrm{d}},\Omega_{\mathrm{e}}\right)  $ who
forever compete with each other for ruling the curvature parameter. Cosmic
expansion is in this case an eternal sequence of transient acceleration (when
DE dominates) and deceleration (when DM dominates) phases. \newline Solutions
such that $\Omega_{\mathrm{d}}+\Omega_{\mathrm{e}}>1$ are unbounded. They
correspond to spatially closed universes, since $\Omega_{k}<0$, in which
cosmic expansion can reverse into contraction at some stage, leading to $H=0$
and consequent singularities in all density parameters. The present formalism
with monotonically growing $\lambda=\ln(a)$ cannot extrapolate beyond in
vanishing $H$ toward cosmic contraction $H<0$, since this would imply
decreasing $\lambda$.

\subsection{Attractiveness of the dark plane \label{sect:att}}

We now turn our attention to the behavior of orbits whose initial conditions
are not in the dark plane but have non vanishing components in $\Omega
_{\mathrm{b}}$ and/or $\Omega_{\mathrm{r}}$. Intuitively one could claim that
these components are going to vanish because the eigenvalues associated to
them are negative, but as the two others, associated to the dark components,
are purely imaginary, the equilibrium is no longer a hyperbolic one hence
Hartmann-Gro$\beta$man theorem says that the linear analysis is not sufficient
to have a complete description of the system behavior. This point is important
and it is often forgotten in the physical literature, it is why we give a
counter example in the appendix \ref{appnonhyp}. However, even if the invariant manifold
methods cannot be straightforwardly used, because the centre manifold is
infinitely flat at $\mathbf{\tilde{x}}^{4}$, we are able, using dynamical
systems tools, to prove the attractiveness of the dark plane, for all orbits
whose initial conditions belong to the hyper-tetrahedron~(\ref{eq:htetra}). A
detailed proof of the latter statement will be provided in the appendix \ref{appattract}.

\subsection{Numerical illustration}

\begin{figure}[th]
\centering
\includegraphics[width=8cm]{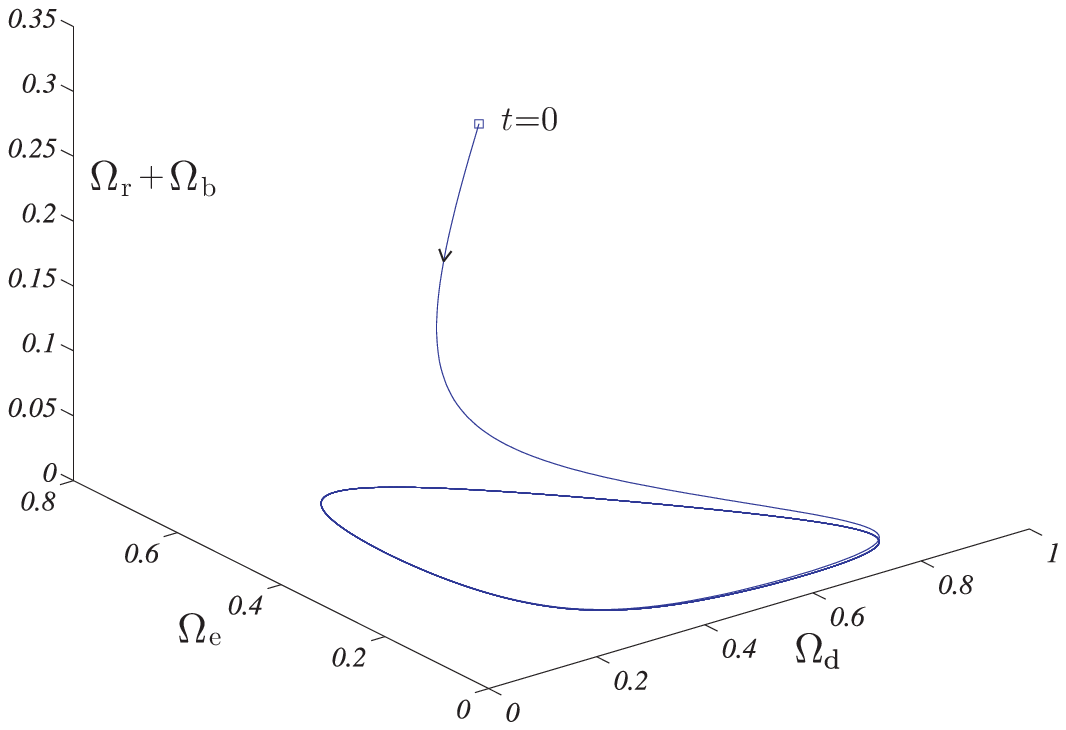}\quad
\includegraphics[width=7cm]{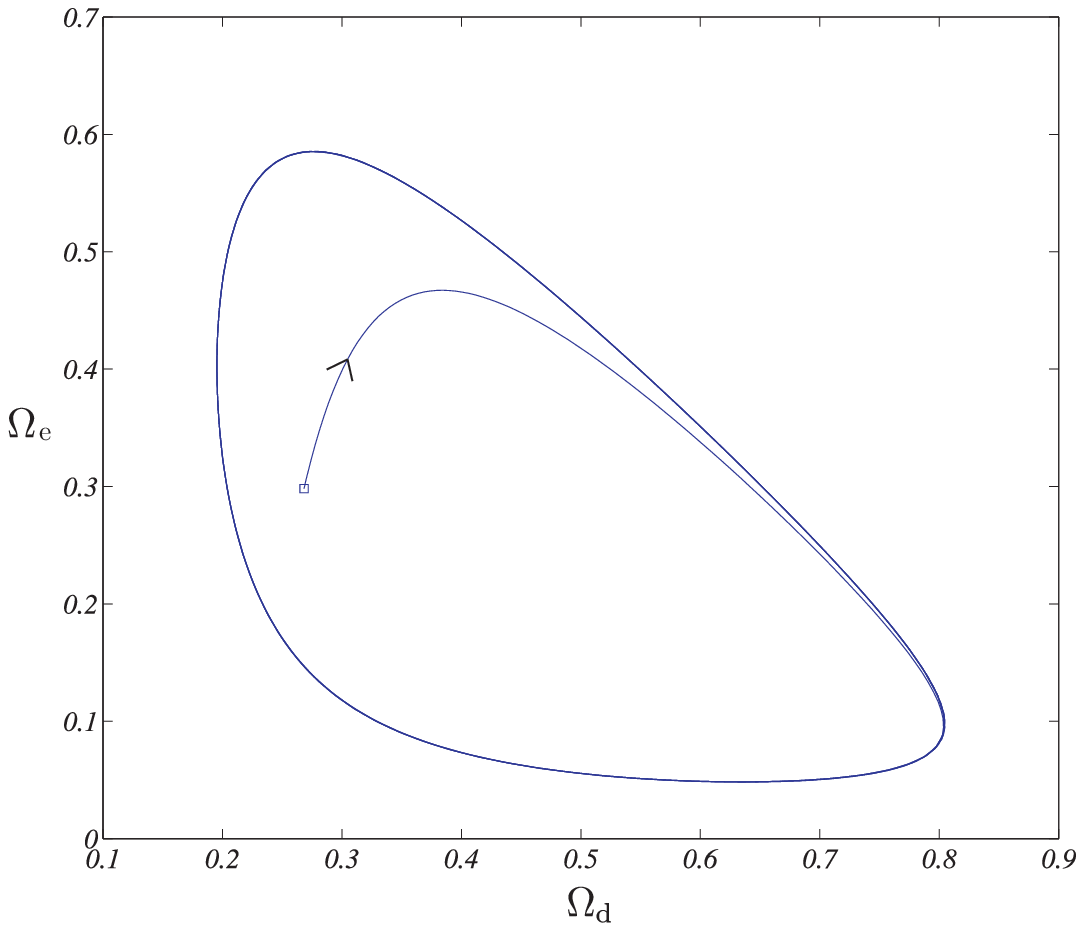}\caption{Time evolution
of the orbit inside $T_{4}$. Left panel: 3D plot of the orbit (the vertical
axis is $\Omega_{\mathrm{r}}(t)+\Omega_{\mathrm{b}}(t)$). Right panel: 2D
projection on the $(\Omega_{\mathrm{d}},\Omega_{\mathrm{e}})$ plane. Parameter
and initial conditions: $\varepsilon=4$, $\Omega_{\mathrm{d}}(0)=0.2683$,
$\Omega_{\mathrm{e}}(0)= 0.2983$, $\Omega_{\mathrm{r}}(0)= 0.0418$ and
$\Omega_{\mathrm{b}}(0)= 0.3008$.}%
\label{fig:3Dplot}%
\end{figure}

As we have obtained a general proof of the attractiveness of the dark plane,
we give only a simple numerical illustration of this fact. We have numerically
solved the dynamical system $\left(  \ln(\mathrm{x})\right)  ^{\prime
}=\mathrm{r}+A\mathrm{x}$ with $\mathbf{x}=\left(  \Omega_{\mathrm{b}}%
,\Omega_{\mathrm{d}},\Omega_{\mathrm{r}},\Omega_{\mathrm{e}}\right)  ^{\top}$,
the community matrix and the capacity vector defined in (\ref{sysdyn1}) with
$\varepsilon=4$. Considering various initial conditions $\mathbf{x}_{0}$ we
always recover an exponential convergence to the dark plane when
$\mathbf{x}_{0}$ has non vanishing first and third components. The figure
\ref{fig:3Dplot} illustrate such a behavior: from the initial condition
$\mathbf{x}_{0}=\left(  0.3008,0.2683,0.0418,0.2983\right)  ^{\top}$, which
belongs to the stable hyper-tetrahedron, we have 3D-plotted the dynamical
evolution of the vector $\left(  \Omega_{\mathrm{d}},\Omega_{\mathrm{e}%
},\Omega_{\mathrm{b}}+\Omega_{\mathrm{r}}\right)  ^{\top}$. As expected the
third component vanishes and the two others are caught by a contour level of
$V_{4}$. View from the top in the right part of the figure \ref{fig:3Dplot} is
particularly explicit about this last fact.

\section{General correspondence between coupled models and Lotka-Volterra
competitive dynamics\label{section4}}

In the previous section, we have deduced the general behavior of two coupled
species in Jungle Universes. We propose to call such cosmological components
twisting species since the special example proposed in the last section
represent an eternal exchange between dark energy and dark matter. We will
illustrate now that such a behavior can be generalized introducing more couplings.

In this section, we extend the previous discussion to a set of $N$
inter-coupled cosmological species and establish the correspondence with
general formulation of competitive Lotka-Volterra models. The goal here is
therefore to rewrite the evolution, with the variable $\lambda=\ln(a)$, of
cosmological density parameters of interacting fluids under the following
Lotka-Volterra form :
\begin{equation}
\mathbf{x}^{\prime}=\mbox{diag}(\mathbf{x}) \mathbf{f}(\mathbf{x}%
)\quad\mbox{with}\,\mathbf{x}\in\mathbb{R}^{n} \label{LV}%
\end{equation}
where $\mbox{diag}(\mathbf{x})$ is the diagonal matrix with $\mathbf{x}$ on
its diagonal, the $i$th component of the vector $\mathbf{x}$ denotes the
population of the $i$th species, $\mathbf{f}(\mathbf{x})=\mathbf{r}%
+A\mathbf{x}$ is the previously defined linear function which combines the
capacity vector $\mathbf{r}$ and the community matrix $A$. Each coupled fluid
characterized by energy density $\rho_{i}$, equation of state parameter
$\omega_{i}$ and obey the following modified conservation equation:
\begin{equation}
\dot{\rho}_{i}+3H\rho_{i}(1+\omega_{i})=\mathcal{Q}_{i}\; ; \; i=1,\cdots,N
\label{cons1}%
\end{equation}
with the energy balance condition imposing that
\begin{equation}
\sum_{i=1}^{N} \mathcal{Q}_{i}=0 \label{balance}%
\end{equation}
where the interaction terms $\mathcal{Q}_{i}$ take the form of a combination
of the involved energy densities:
\begin{equation}
\mathcal{Q}_{i}=\sum_{j=1}^{N}\beta_{ij}\rho_{j}\cdot
\end{equation}
Defining the density parameters $\Omega_{i}=\frac{8\pi G\rho_{i}}{3H^{2}}$,
and recalling that the deceleration parameter can be written as
\[
q=-\frac{\ddot{a}a}{\dot{a}^{2}}=\frac{1}{2}\sum_{i=1}^{N} \Omega_{i}
(1+3\omega_{i})
\]
then Eq. (\ref{cons1}) becomes
\begin{equation}
\dot{\Omega_{i}}=\frac{8\pi G \mathcal{Q}_{i}}{3H^{2}}+H\Omega_{i}\left(
2-3(1+\omega_{i})+\sum_{j=1}^{N} \Omega_{j} (1+3\omega_{j})\right)
\cdot\label{cons2}%
\end{equation}
To rewrite the above equation under Lotka-Volterra form, it is then mandatory
to set
\begin{equation}
\mathcal{Q}_{i}=\sum_{j=1}^{N} \beta_{ij}\rho_{j}\equiv H\Omega_{i}\sum
_{j=1}^{N} \varepsilon_{ij}\rho_{j} \label{Qi}%
\end{equation}
or, equivalently that the coefficients $\beta_{ij}$ are no longer constant but
are given by
\[
\beta_{ij}=H\Omega_{i} \varepsilon_{ij}
\]
with $\varepsilon_{ij}$ arbitrary parameters to be specified further.
Lotka-Volterra dynamics therefore requires non-linear interaction terms. Given
Eq. (\ref{Qi}), one can directly rewrite Eq.(\ref{cons2}) under Lotka-Volterra
form (\ref{LV}) with the following glossary:
\begin{align}
x_{i}  &  =\Omega_{i}\nonumber\\
(\cdot)^{\prime}  &  =\frac{d(\cdot)}{d\ln(a)}\nonumber\\
r_{i}  &  =-(1+3\omega_{i})\label{gloss}\\
A_{ij}  &  =1+3\omega_{j}+\varepsilon_{ij}\nonumber
\end{align}
The energy balance constraint Eq.(\ref{balance}) with the hypothesis
(\ref{Qi}) now reduces to
\begin{equation}
\sum_{i=1}^{N} \Omega_{i} \left(  \sum_{j=1}^{N} \varepsilon_{ij}\Omega_{j}
\right)  =0
\end{equation}
which imposes that the interaction parameters $\varepsilon_{ij}$ are
\textit{antisymmetric}:
\[
\varepsilon_{ij}=-\varepsilon_{ji} \; ; \; \varepsilon_{ii}=0\cdot
\]
In the context of cosmology we find solution of this ODE system in the
hyper-tetraedron
\[
\mathcal{T}=\left\{  1>\sum_{i=1}^{N} x_{1}\right\}  \bigcap_{i=1}^{N}
\left\{  x_{i}>0\right\}
\]
The generalization of the results obtained in the previous section show that
ODE system (\ref{LV}) has generically a lot of equilibria but we are
interested only by the ones who haven't vanishing component (i.e. the ones not
lying on an axis). These "interesting" equilibria are $\tilde{\mathbf{x}}$
such that $A\tilde{\mathbf{x}}+\mathbf{r}=\mathbf{0}$. We can now apply this
general formulation to the case of several interacting species.


\subsection{Two species in interaction}

This case $N=2$ has been treated in details in section 3 for specific values
of the equation of state parameters $(\omega_{1},\omega_{2})=(0,-1)$ and
serves here as a validation of the glossary (\ref{gloss}). Setting
$\varepsilon_{12}=-\varepsilon_{21}\equiv\varepsilon$ the unique non-vanishing
component of the interaction tensor $\varepsilon_{ij}$, we obtain after some
computation the following equilibria of the cosmological Lotka-Volterra
system
\begin{align}
\Omega_{1}^{eq}  &  =-\frac{3\omega_{2}+1}{\varepsilon}\\
\Omega_{2}^{eq}  &  =+\frac{3\omega_{1}+1}{\varepsilon}%
\end{align}
These equilibria are density parameters in open universes ($\Omega_{k}<1$) and
then must satisfy $0<\Omega_{i}^{eq}<1$. This condition constrains the choice
of $\varepsilon$ once the choice of the nature of the interacting fluids has
been chosen by fixing $\omega_{2}$ and $\omega_{1}$.

\subsection{The interplay between three coupled species : Jungle triads}

Let us set $\varepsilon_{12}=e_{1}$, $\varepsilon_{13}=e_{2}$ and
$\varepsilon_{23}=e_{3}$ and compute the corresponding equilibria to find
\begin{align}
\Omega_{1}^{eq}  &  =+\frac{e_{3}-3\omega_{2}+3\omega_{3}}{e_{1}-e_{2}+e_{3}%
}\nonumber\\
\Omega_{2}^{eq}  &  =-\frac{e_{2}-3\omega_{1}+3\omega_{3}}{e_{1}-e_{2}+e_{3}%
}\\
\Omega_{3}^{eq}  &  =+\frac{e_{1}-3\omega_{1}+3\omega_{2}}{e_{1}-e_{2}+e_{3}%
}\nonumber
\end{align}

\begin{figure}[th]
\begin{center}
\includegraphics[scale=1]{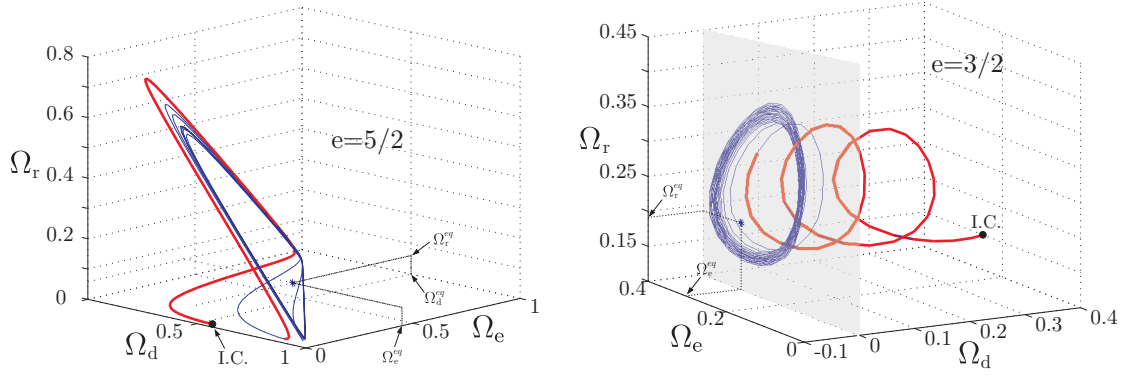}
\end{center}
\caption{Evolution of the three coupled density parameters, in the 3D phase
space. The beginning of the orbit is overlined. Initial condition is indicated
by a black dot. Relevant equilibria are indicated by a star.}%
\label{triades}%
\end{figure}

Let us remark that in all cases of fluids and coupling we have $\sum_{i=1}%
^{3}\Omega_{i}^{eq}=1$. This fact seems generic for odd values of the number
of interacting fluids. If we now impose the fact that the density parameters
are comprised between $0$ and $1$ ($0<\Omega_{i}^{eq}<1$) the constraint on
interaction parameters $e_{1},e_{2},e_{3}$ is very complicated, but allows a
lot of possibilities. Let us illustrate this case with an example. We consider
that the three fluids are made of (1) non-relativistic matter $\omega_{1}=0$,
($x_{1}=\Omega_{\mathrm{d}}$); (2) dark energy $\omega_{2}=-1$, ($x_{2}%
=\Omega_{\mathrm{e}}$) and (3) some relativistic particles $\omega_{3}=1/3$,
($x_{3}=\Omega_{\mathrm{r}}$) all coupled with interaction parameters
$e_{1}=e_{2}=e$ and $e_{3}=\varepsilon$. The corresponding equilibria are
\[
\Omega_{\mathrm{d}}^{eq}=\frac{4+\varepsilon}{\varepsilon},\quad
\Omega_{\mathrm{e}}^{eq}=-\frac{1+e}{\varepsilon} \quad\mbox{and}\;
\Omega_{\mathrm{r}}^{eq}=\frac{e-3}{\varepsilon}%
\]
Providing $\varepsilon<-4$ and $e\in[-1,3]$ equilibria are cosmologically
acceptable. Choosing for example $\varepsilon=-8$, the spectrum of the
jacobian matrix near the equilibrium is composed by a real number
$\lambda=1-\frac{e}{2}$ and two purely imaginary and complex conjugated
numbers $\lambda_{\pm}=\pm\frac{i}{2}\sqrt{ 2\left|  (e+1)(e-3)\right|  }$.
When $e\in[-1,2]$, as $\lambda>0$ the system twists \emph{outward}
$(0,\Omega_{\mathrm{e}}^{eq},\Omega_{\mathrm{r}}^{eq})$ staying in the
corresponding 3-tetraedron, collapsing on the $\Omega_{\mathrm{d}}=0$ plane.
When $e\in[2,3]$, as $\lambda<0$ the system twists \emph{toward} a limit cycle
contained in a plane of non vanishing density and including the equilibrium.
These results are illustrated on the figure \ref{triades}. 

\subsection{Jungle quartets}

\begin{figure}[th]
\begin{center}
\includegraphics[width=\linewidth]{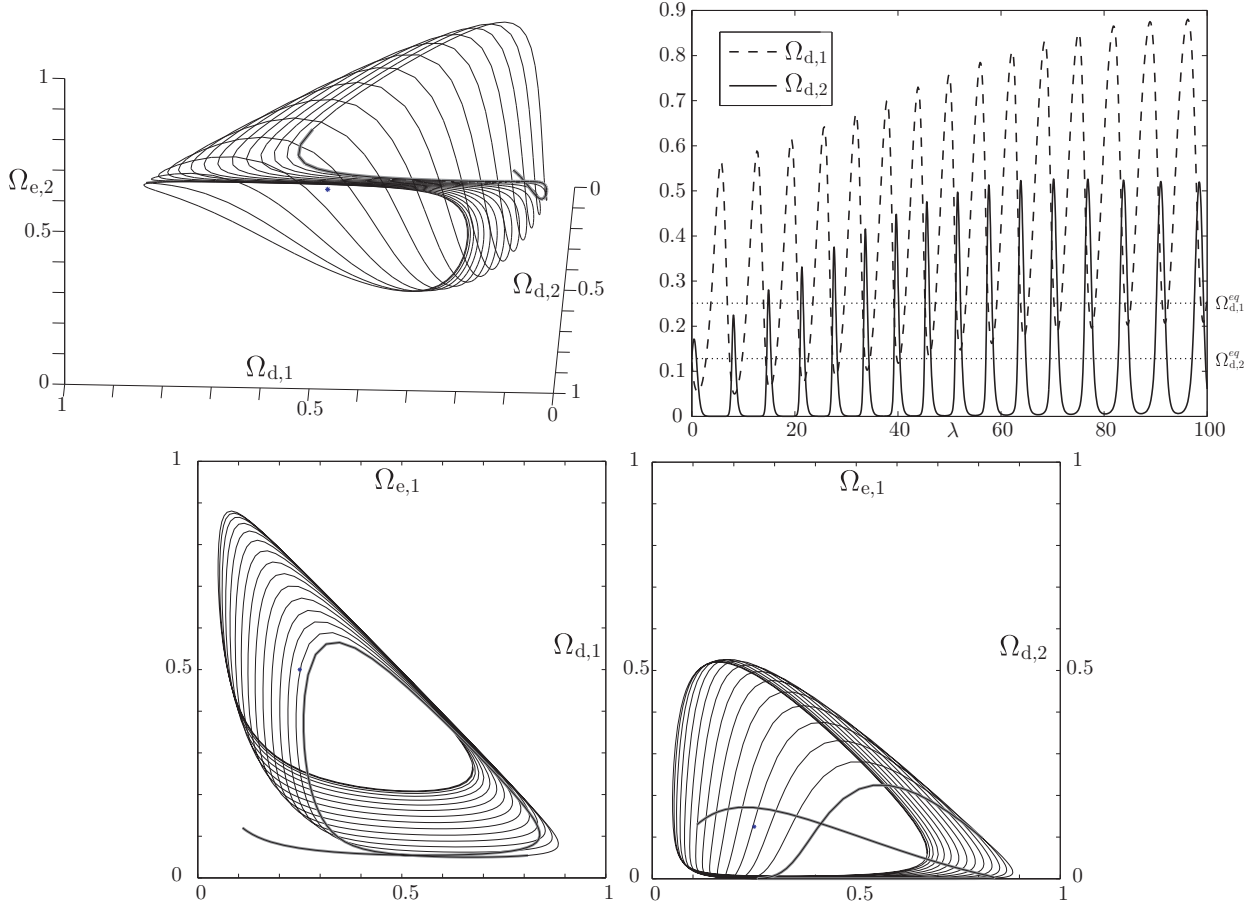} .
\end{center}
\caption{Jungle quartet : The left top panel is a 3D section of the 4D phase
space. The right top panel is a representation of $\Omega_{\mathrm{d,1}}%
^{eq}(\lambda)$ and $\Omega_{\mathrm{d,2}}^{eq}(\lambda)$, the corresponding
equilibria are indicated by doted horizontal lines. The two bottom panels are
2D sections of the 4D phase space. For the phase space sections, the beginning
of the orbits are overlined and the relevant equilibria are indicated by a
star. Initial conditions for the numerical integration are $x_{1}(0)=0.11$,
$x_{2}(0)=0.12$, $x_{3}(0)=0.13$ and $x_{4}(0)=0.14$}%
\label{quatuor}%
\end{figure}

With $N=4$, the number of free parameters in the scheme (10 in total with 6
for interactions and 4 for equations of state) is too high to be fully
constrained by requirements of positiveness and boundedness of density
parameters for instance. As for $N=2$, the positions of the equilibria once
again depend on all parameters. If we set $\varepsilon_{12}=e_{1}$,
$\varepsilon_{13}=e_{2}$, $\varepsilon_{14}=e_{3}$, $\varepsilon_{23}=e_{4}$,
$\varepsilon_{24}=e_{5}$ and $\varepsilon_{34}=e_{6}$, we find that the
positions of the equilibria are given by
\begin{align}
\Omega_{1}^{eq}  &  =-\frac{e_{4}-e_{5}+e_{6}+3(e_{4}\omega_{4}-e_{5}%
\omega_{3}+e_{6}\omega_{2})}{e_{1}e_{6}-e_{2}e_{5}+e_{4}e_{3}}\nonumber\\
\Omega_{2}^{eq}  &  = +\frac{e_{2}-e_{3}+e_{6}+3(e_{2}\omega_{4}-\omega
_{3}e_{3}+e_{6}\omega_{1})}{e_{1}e_{6}-e_{2}e_{5}+e_{4}e_{3}}\nonumber\\
\Omega_{3}^{eq}  &  =-\frac{e_{1}-e_{3}+e_{5}+3(e_{1}\omega_{4}+\omega
_{1}e_{5}-\omega_{2}e_{3})}{e_{1}e_{6}-e_{2}e_{5}+e_{4}e_{3}}\\
\Omega_{4}^{eq}  &  = +\frac{e_{1}-e_{2}+e_{4}+3(e_{1}\omega_{3}-e_{2}%
\omega_{2}+e_{4}\omega_{1})}{e_{1}e_{6}-e_{2}e_{5}+e_{4}e_{3}}\nonumber
\end{align}
Since this system of 4 cosmological coupled species is equivalent to 4D
Lotka-Volterra system, chaos can emerge \cite{wang} for specific choices of
parameters in a so-called normal system where all $r_{i}$ are positive, which
means among cosmological fluids with $\omega_{i}<-1/3$. As an illustration we
propose a double twist in a universe filled by two kinds of dark energy and
two kinds of dark matter all interacting. We choose $\omega_{1}=-1$,
($x_{1}=\Omega_{\mathrm{e,1}}$); $\omega_{2}=0$, ($x_{2}=\Omega_{\mathrm{d,1}%
}$); $\omega_{3}=0$, ($x_{3}=\Omega_{\mathrm{d,2}}$) and $\omega_{4}=-1$,
($x_{4}=\Omega_{\mathrm{e,2}}$) for the fluid components, and $e_{1}=-4$,
$e_{2}=1$, $e_{3}=-2$, $e_{4}=-1/2$, $e_{5}=1$ and $e_{6}=\varepsilon$, we get
the following equilibria
\[
\Omega_{\mathrm{e,1}}^{eq}=\frac{1}{4},\;\Omega_{\mathrm{d,1}}^{eq}=\frac
{1}{2},\;\Omega_{\mathrm{d,2}}^{eq}=\frac{2}{\varepsilon} ,\;\Omega
_{\mathrm{e,2}}^{eq}=\frac{1}{\varepsilon}
\]
The condition on the density parameters then gives $\varepsilon>12$. Taking
$\varepsilon=16$ we get four complicated but, purely imaginary and conjugated
eigenvalues for the Jacobian matrix around the equilibrium:
\[
\lambda_{1}^{\pm}=\pm i\frac{\sqrt{51134+6\sqrt{69956601}}}{262}
\quad\mbox{and}\quad\lambda_{2}^{\pm}=\pm i\frac{\sqrt{51134-6\sqrt{69956601}%
}}{262}
\]
The corresponding dynamics is the double twist plotted on figure
\ref{quatuor}. 

\section{Conclusion\label{section5}}

Let us summarize the main points obtained in this paper :

\begin{itemize}
\item We have formulated the classical dynamics of Friedmann Universes in the
context of the generalized Lotka-Volterra equation. Without coupling, this
formulation allows a very simple and pedagogic interpretation of the evolution
of these universes. Varying parameters describing the nature of the fluids one
can easily understand the corresponding behavior of the so-called Jungle universes.

\item Cyclic behaviors has been speculated by Lip\cite{4} and Arevalo et al.\cite{5} when FL universe contains exotic "phantom dark matter" fluids (with barotropic index  $\omega<-1$) coupled with dark matter; using the generalized Lotka-Volterra formulation of the coupled FL universe we have obtained a general Lyapunov function in the context of the standart cosmological model. This function allows us to rigorously prove the existence of cyclic behavior of FL universe when standart fluids (with barotropic index $\omega>-\frac{1}{3}$) are coupled to dark energy (with barotropic index $\omega<-\frac{1}{3}$). 

\item In the case of 3 or 4 interacting batrotropic fluids, we have found
more  complex cyclic behavior of the universe : an expanding twist for $N=3$ and a double twist for $N=4$.

\item Following the results of the population dynamics, we conjecture that
chaos occurs as a rule for the dynamics of universes filled by more than 3
interacting fluids.
\end{itemize}

We conclude by claiming that the presented analogy with Lotka-Volterra
dynamical systems has offered new unexpected and interesting applications to
coupled models in cosmology. Twisting species naturally produce transient
phenomena in cosmic expansion, an original feature that could make cosmic
coincidence a non unique and therefore less problematic feature.

\subsubsection*{Acknowledgement}

J.P. thanks Fr\'{e}d\'{e}ric Jean for helpful discussions.
\appendix

\section{Counter example : a linear center which is actually nonlinearly
unstable \label{appnonhyp}}

Consider the dynamical system
\[
\left\{
\begin{array}
[c]{l}%
\dot{x}=-y+x^{3}\\
\dot{y}=x+y^{3}%
\end{array}
\right.
\]
The origin is an equilibrium, the eigenvalues of the jacobian near this
equilibrium are $\pm i$. In the linear approximation the origin seems to be a
center. But as the eigenvalues have no real part, the system is not hyperbolic
and almost nothing can be assumed about the non linear dynamics considering
only the linear one around the equilibrium. In this particular case one can
prove that the origin is a repulsive focus, and it is then actually unstable.
As a matter of fact, considering the intersection $M$ between an orbit and the
the circle $x^{2}+y^{2}=R^{2}$. The angle $\alpha$ between the tangent in $M$
to the orbit and and the tangent in $M$ to the circle is given for any radius
$R$ by the expression
\[
\cos\alpha=\left(  \dot{x},\dot{y}\right)  \cdot\left(  2x,2y\right)  =\left(
-y+x^{3},x+y^{3}\right)  \cdot\left(  2x,2y\right)  =2x^{4}+2y^{4}>0
\]
Hence, each orbit is going out from any circle of radius $R>0$ and the origin
is unstable.

\section{Proof of the stability of the $\Omega_{\mathrm{d}}-\Omega
_{\mathrm{e}}$ plane \label{appattract}}

The aim of this section is to provide a simple proof of the attractiveness of
the $\Omega_{\mathrm{d}}-\Omega_{\mathrm{e}}$ plane for all orbits whose
initial conditions belong to the hyper-tetrahedron:
\[
T_{4}=\{\Omega_{\mathrm{d}}>0\}\cup\{\Omega_{\mathrm{e}}>0\}\cup
\{\Omega_{\mathrm{r}}>0\}\cup\{\Omega_{\mathrm{b}}>0\}\cup\{\Omega
_{\mathrm{d}}+\Omega_{\mathrm{e}}+\Omega_{\mathrm{r}}+\Omega_{\mathrm{b}%
}<1\}\, .
\]

Let us recall the ODE system describing the equations of motion:
\begin{equation}
\left\{
\begin{array}
[c]{l}%
\Omega_{\mathrm{d}}^{\prime}=\Omega_{\mathrm{d}}\left[  \Omega_{\mathrm{d}%
}+(\varepsilon+1+3\omega_{\mathrm{e}})\Omega_{\mathrm{e}}+2\Omega_{\mathrm{r}%
}+\Omega_{\mathrm{b}}-1\right] \\
\Omega_{\mathrm{e}}^{\prime}=\Omega_{\mathrm{e}}\left[  (1-\varepsilon
)\Omega_{\mathrm{d}}+\left(  1+3\omega_{\mathrm{e}}\right)  \Omega
_{\mathrm{e}}+2\Omega_{\mathrm{r}}+\Omega_{\mathrm{b}}-1-3\omega_{\mathrm{e}%
}\right] \\
\Omega_{\mathrm{r}}^{\prime}=\Omega_{\mathrm{r}}\left[  \Omega_{\mathrm{d}%
}+\left(  1+3\omega_{\mathrm{e}}\right)  \Omega_{\mathrm{e}}+2\Omega
_{\mathrm{r}}+\Omega_{\mathrm{b}}-2\right] \\
\Omega_{\mathrm{b}}^{\prime}=\Omega_{\mathrm{b}}\left[  \Omega_{\mathrm{d}%
}+\left(  1+3\omega_{\mathrm{e}}\right)  \Omega_{\mathrm{e}}+2\Omega
_{\mathrm{r}}+\Omega_{\mathrm{b}}-1\right]
\end{array}
\right.  \label{eq:ode}%
\end{equation}

To prove our claim we need to prove first the invariance with respect to the
flow(\ref{eq:ode}) of the hyper-tetrahedron $T_{4}$. The invariance of each
coordinates hyperplanes is trivial and follows straightforwardly
from~(\ref{eq:ode}). For instance any solution such that $\Omega_{\mathrm{d}%
}(0)=0$ will have $\Omega_{\mathrm{d}}(\lambda)=0$ for all $\lambda$, then
using the uniqueness of the Cauchy problem we can ensure that any solution
with $\Omega_{\mathrm{d}}(0)>0$ will never cross the hyperplane $\Omega
_{\mathrm{d}}=0$. A very similar analysis can be performed for the remaining cases.

Let us now consider the remaining piece of the boundary of $T_{4}$, that is
the hyperplane$\{\Omega_{\mathrm{d}}+\Omega_{\mathrm{e}}+\Omega_{\mathrm{r}%
}+\Omega_{\mathrm{b}}=1\}$. A straightforward computation gives:
\[
\left(  \Omega_{\mathrm{d}}+\Omega_{\mathrm{e}}+\Omega_{\mathrm{r}}%
+\Omega_{\mathrm{b}}\right)  ^{\prime}=\left[  \Omega_{\mathrm{d}}%
+(1+3\omega_{\mathrm{e}})\Omega_{\mathrm{e}}+2\Omega_{\mathrm{r}}%
+\Omega_{\mathrm{b}}\right]  \left[  \Omega_{\mathrm{d}}+\Omega_{\mathrm{e}%
}+\Omega_{\mathrm{r}}+\Omega_{\mathrm{b}}-1\right]  \,,
\]
thus any solution with initial conditions
\[
\Omega_{\mathrm{d}}(0)+\Omega_{\mathrm{e}}(0)+\Omega_{\mathrm{r}}%
(0)+\Omega_{\mathrm{b}}(0)=1\,,
\]
will always satisfies the constraint
\[
\Omega_{\mathrm{d}}(\lambda)+\Omega_{\mathrm{e}}(\lambda)+\Omega_{\mathrm{r}%
}(\lambda)+\Omega_{\mathrm{b}}(\lambda)=1\quad\forall\lambda\,.
\]
Thus once again the uniqueness result of the Cauchy problem implies that any
solution such that $\Omega_{\mathrm{d}}(0)+\Omega_{\mathrm{e}}(0)+\Omega
_{\mathrm{r}}(0)+\Omega_{\mathrm{b}}(0)<1$, will never reach the hyperplane
$\Omega_{\mathrm{d}}+\Omega_{\mathrm{e}}+\Omega_{\mathrm{r}}+\Omega
_{\mathrm{b}}=1$.

Finally putting together the above partial results, we can conclude that any
orbit with initial condition inside $T_{4}$ will never leave it.

A by-product of the invariance of the tetrahedron is that orbits inside
$T_{4}$ will always have positive projections on the axes. This allows us to
compute the distance from the plane $\left(  \Omega_{\mathrm{d}}%
,\Omega_{\mathrm{e}}\right)  $ using the linear function $F(\Omega
_{\mathrm{r}},\Omega_{\mathrm{b}})=\Omega_{\mathrm{r}}+\Omega_{\mathrm{b}}$,
which is zero if and only if $\Omega_{\mathrm{r}}=\Omega_{\mathrm{b}}=0$, that
is the point belongs to the plane $\left(  \Omega_{\mathrm{d}},\Omega
_{\mathrm{e}}\right)  $.

We can then compute the Lie derivative of $F$ and prove that its restriction
to $T_{4}$ is strictly negative, hence $F(\Omega_{\mathrm{r}}(\lambda
),\Omega_{\mathrm{b}}(\lambda))\rightarrow0$ for $\lambda\rightarrow+\infty$
and because of the positiveness of $\Omega_{\mathrm{r}}\left(  \lambda\right)
$ and $\Omega_{\mathrm{b}}\left(  \lambda\right)  $ we can conclude that both
$\Omega_{\mathrm{r}}\left(  \lambda\right)  $ and $\Omega_{\mathrm{b}}\left(
\lambda\right)  $ goes asymptotically to zero.

To prove the latter claim let us compute the derivative of $F$ along the flow
of~(\ref{eq:ode}):
\[
\left.  \frac{dF}{dt}\right\vert _{\mathrm{flow}}=\left[  \Omega_{\mathrm{d}%
}+(1+3\omega_{\mathrm{e}})\Omega_{\mathrm{e}}+2\Omega_{\mathrm{r}}%
+\Omega_{\mathrm{b}}-1\right]  \left[  \Omega_{\mathrm{r}}+\Omega_{\mathrm{b}%
}\right]  -\Omega_{\mathrm{r}}\,,
\]
because of our previous result $\Omega_{\mathrm{d}}(\lambda)+\Omega
_{\mathrm{b}}(\lambda)-1<-\Omega_{\mathrm{e}}(\lambda)-\Omega_{\mathrm{r}%
}(\lambda)$ for all $\lambda$.\ Using this inequality we get
\[
\left.  \frac{dF}{dt}\right\vert _{\mathrm{flow}}<\left(  3\omega_{\mathrm{e}%
}\Omega_{\mathrm{e}}+\Omega_{\mathrm{r}}\right)  \left(  \Omega_{\mathrm{r}%
}+\Omega_{\mathrm{b}}\right)  -\Omega_{\mathrm{r}}\,=3\omega_{\mathrm{e}%
}\Omega_{\mathrm{e}}\left(  \Omega_{\mathrm{r}}+\Omega_{\mathrm{b}}\right)
+\left(  \Omega_{\mathrm{r}}+\Omega_{\mathrm{b}}-1\,\right)  \Omega
_{\mathrm{r}}%
\]
let us observe that the right hand side is strictly negative, in fact
\[
3\omega_{\mathrm{e}}\Omega_{\mathrm{e}}(\Omega_{\mathrm{r}}+\Omega
_{\mathrm{b}})<0
\]
provided and $\Omega_{\mathrm{e}}$ is associated to the dark energy $\left(
\omega_{\mathrm{e}}<-\frac{1}{3}<0\right)  $ and
\[
\Omega_{\mathrm{r}}+\Omega_{\mathrm{b}}-1<\Omega_{\mathrm{d}}+\Omega
_{\mathrm{e}}+\Omega_{\mathrm{r}}+\Omega_{\mathrm{b}}-1<0.
\]

This concludes our proof of the attractiveness of the dark plane for all
orbits whose initial conditions belong to $T_{4}$.

\setcounter{section}{1}

\end{document}